\documentclass[cameraready]{Interspeech}

\title{Towards Robust Speech Deepfake Detection via Human-Inspired Reasoning}

\author[affiliation={1}]{Artem}{Dvirniak}
\author[affiliation={2,3,4}]{Evgeny}{Kushnir}
\author[affiliation={3,5}]{Dmitrii}{Tarasov}
\author[affiliation={6}]{Artem}{Iudin}
\author[affiliation={7}]{Oleg}{Kiriukhin}
\author[affiliation={2,8}]{Mikhail}{Pautov}
\author[affiliation={2,4,6},correspondingauthor]{Dmitrii}{Korzh}
\author[affiliation={2,4,6}]{Oleg Y.}{Rogov}

\address{
    $^1$MIRAI,
    $^2$AXXX,
    $^3$HSE,
    $^4$Applied AI Institute,
    $^5$Fusion Brain Lab, AXXX, \\
    $^6$MTUCI,
    $^7$City University of Hong Kong,
    $^8$Trusted AI Research Center, RAS
}

\email{d.s.korzh@mtuci.ru}


\usepackage{comment}

\usepackage{graphicx} 
\usepackage{xcolor}
\usepackage{amsfonts}
\usepackage{dsfont}
\usepackage{booktabs}
\usepackage{makecell}
\usepackage{multirow}
\usepackage{algorithm}
\usepackage{algpseudocode} %
\usepackage{amsmath,amssymb}
\usepackage{listings}

\begin{document}

\maketitle

\begin{abstract}
    The modern generative audio models can be used by an adversary in an unlawful manner, specifically, to impersonate other people to gain access to private information. To mitigate this issue, speech deepfake detection (SDD) methods started to evolve. Unfortunately, current SDD methods generally suffer from the lack of generalization to new audio domains and generators. More than that, they lack interpretability, especially human-like reasoning that would naturally explain the attribution of a given audio to the bona fide or spoof class and provide human-perceptible cues. In this paper, we propose HIR-SDD, a novel SDD framework that combines the strengths of Large Audio Language Models (LALMs) with the chain-of-thought reasoning derived from the novel proposed human-annotated dataset. Experimental evaluation demonstrates both the effectiveness of the proposed method and its ability to provide reasonable justifications for predictions.
\end{abstract}

\section{Introduction}

Contests, such as ASVspoof~\cite{todisco2019asvspoof,wang2024asvspoof}, ADD~\cite{yi2022add, yi2023add}, and Singfake~\cite{zang2024singfake,zhang2024svdd} drive the progress in audio and speech deepfake detection (SDD) research, providing high-quality deepfake data and fair evaluation protocols. SDD and spoofing-aware speaker verification (SASV) research primarily focuses on the architecture design, including task-specific front-ends~\cite{tak2021end}, graph-attention networks~\cite{jung2022aasist}, self-supervised (SSL) audio encoders~\cite{tak2022automatic, aliyev2024intema}, and architecture modifications~\cite{zhang2024improving, borodin2024aasist3} to improve the empirical performance of the models. Additionally, augmentation strategies ~\cite{tak2022rawboost}, optimizer choices~\cite{foret2020sharpness},  and representation learning-based approaches~\cite{ding2023samo, kim2025enhancing, yang2025generalizable} are also investigated in SDD research.

Unfortunately, SDD remains challenging due to distribution shifts across spoofing methods, speech domains, and transformations, as evidenced by results from the aforementioned contests. A common strategy to improve empirical performance is to increase training diversity and model capacity. However, SDD systems still fail to generalize to unseen domains, highlighting the need for more robust and explainable detection approaches. Such tools are particularly important for risk-sensitive applications, such as biometrics and banking, yet this direction remains underexplored in SDD research.

Recently, Large Language Models (LLMs)~\cite{achiam2023gpt, guo2025deepseek} and Large Audio Language Models (LALMs), such as SALMONN~\cite{tang2023salmonn}, Qwen2-Audio~\cite{chu2024qwen2}, and Audio-Flamingo~3~\cite{goel2025audio}, have demonstrated strong reasoning capabilities. Chain-of-thought (CoT)~\cite{wei2022chain} and related reasoning methods~\cite{guo2025deepseek} often improve performance by extracting intermediate rationales. Although CoT mainly acts as an empirical mechanism and does not necessarily reflect human reasoning, careful training and grounding can produce explanations that are consistent with the input and useful for analysis. Similar ideas have been explored for audio question answering and captioning (including music understanding)~\cite{yang2025sakura, tian2025step}, disease detection from speech~\cite{park2025reasoning}, emotion recognition~\cite{mai2025chain}, and interpretable audio quality assessment and hard-label SDD~\cite{wang2025speechllm}. 

However, reasoning-based approaches remain relatively uncommon in SDD. 
One practical limitation is the lack of open-source datasets with high-quality human explanations for training and evaluation. Moreover, existing LALMs, even state-of-the-art, drastically underperform in zero- and few-shot SDD, thereby limiting the reliability of their reasoning traces~\cite {gu2025allm4add}.
To address these gaps, we introduce a human-annotated dataset for CoT training and evaluation of SDD models and propose \texttt{HIR-SDD}, a human-inspired reasoning framework that combines hard-label and CoT-supervised fine-tuning, grounding, and reinforcement learning (RL) methods. 
\\
\textbf{Our contributions are summarized as follows:}
\begin{itemize}
    \item We present\footnote{\url{https://hf.co/datasets/marsianin500/HIR-SDD}} a new dataset of human-annotated reasoning traces for 41k bona fide and spoof speech samples partially curated from existing open-source datasets.
    \item We propose a hard-label and CoT pipelines that achieve both strong countermeasure and reasoning explainability performance. The source code is available online\footnote{\url{https://github.com/dkorzh10/HIR-SDD}}.
    \item We provide further improvement and evaluation strategies of reasoning-capable SDD models. 
\end{itemize}

\section{Related work}

ALLM4ADD~\cite{gu2025allm4add} is among the first studies to explore large-scale LALMs for SDD. The authors report weak zero-shot performance for Qwen and Qwen2-Audio~\cite{chu2024qwen2}, and then fine-tune the models with a simple prompt requesting a binary ``Real''/``Fake'' response for the input speech. Across several benchmarks, the resulting systems match or outperform smaller state-of-the-art detectors. The study also analyzes training data scaling, sensitivity to LoRA~\cite{hu2022lora}  rank, and prompt design. However, the evaluation in~\cite{gu2025allm4add} does not include the more recent ASVspoof~5 benchmark~\cite{wang2024asvspoof}. Moreover, the analysis focuses mainly on hard-label classification, leaving the reasoning and interpretability capabilities of LALMs largely unexplored. 

In~\cite{nguyen2026analyzing}, the authors study the robustness of SDD models that produce reasoning traces alongside binary decisions. Explanations are evaluated along three aspects: perception quality (e.g., captioning or recognition fidelity), coherence between reasoning and prediction, and robustness under domain shift caused by audio transformations and adversarial perturbations~\cite{szegedy2013intriguing}. The results indicate that reasoning traces do not consistently improve detection accuracy: well-grounded models may benefit, while other LALMs degrade due to hallucinated justifications. The study mainly audits explanation sensitivity under perturbations rather than improving standard SDD evaluation protocols. Experiments are also limited to older generative conditions from ASVspoof~19.
Step-Audio-R1~\cite{tian2025step} addresses audio reasoning issues by grounding explanations to real acoustic evidence rather than textual hallucinations. The approach combines CoT-style supervised fine-tuning (SFT) and RLVR~\cite{wen2025reinforcement}, followed by Modality-Grounded Reasoning Distillation (MGRD). MGRD iteratively self-distills partially correct CoT traces while excluding batches where traces are fully correct or almost entirely incorrect. This process improves performance across speech, environmental sound, and music tasks. However, the method has not been applied to SDD.

Explainable AI methods for audio are surveyed in~\cite{akman2024audio}. In~\cite{li2024interpretable},  an interpretability method for SDD by modifying an SSL-based detector and introducing a class-activation-style representation to highlight influential time regions is proposed. However, evaluation is limited to ASVspoof~2019, and does not provide textual explanations. In~\cite{bolanos2025benchmarking}, a benchmark for time-localized, model-agnostic post-hoc explanations in audio classification is introduced, using temporally annotated events as proxy ground truth. HoliAntiSpoof~\cite{xu2026holiantispoof} applies LALMs to jointly reason about spoofing mechanisms, temporal localization, and semantic artifacts. While it improves hard-label performance and interpretability compared to binary classifiers, it does not explicitly evaluate the quality of SDD reasoning, requires specialized re-training and does not provide results on ASVspoof5.

It is worth mentioning that humans also struggle with SDD~\cite{barrington2025people}. The study~\cite{warren2024better} evaluates responses from $1200$ annotators and reports $73\%$ binary accuracy on detecting deepfakes from ASVspoof~21~\cite{yamagishi2021asvspoof}, WaveFake~\cite{frank2021wavefake}, and FakeAVCeleb~\cite{khalid2021fakeavceleb}. Annotators also provide free-form explanations, later grouped into eight cue categories (e.g., prosody, liveness, quality). The authors compare these cues with automated detectors to analyze when humans outperform machines and which cues are reliable. Although the dataset is diverse, the public release includes only keywords rather than full explanations and does not clearly separate binary decisions from reasoning comments.

\section{Dataset collection}


\subsection{Audio sources}
\label{sec:audio_sources}
Audio was primarily collected from several open-source SDD datasets in English and Russian, including ASVspoof~5~\cite{wang2024asvspoof}, PyAra~\cite{efanov2024comparison}, LibriSecVoc~\cite{sun2023ai}, MLAAD~\cite{muller2024mlaad}, DFADD~\cite{du2024dfadd}, and M-AILABS~\cite{MAILABS_2017}. We also included bona fide speech from Golos~\cite{karpov2021golos}, SOVA~\cite{sova_dataset}, and Russian LibriSpeech (RuLS)~\cite{ruls}. In addition, we incorporated a subset of annotated samples from the SpeechEval dataset introduced in SpeechLLM-as-Judges~\cite{wang2025speechllm}. We further generated audio using the open-source XTTS-V2~\cite{casanova2024xtts} model and several ESpeech~\cite{espeech_tts_rlv2} models. To diversify and strengthen the test set, we also synthesized several thousand samples using proprietary ElevenLabs~\cite{elevenlabs_tts_api} models, approximating real-world conditions.
\begin{figure}[tbh]
  \centering
  \includegraphics[width=\linewidth]{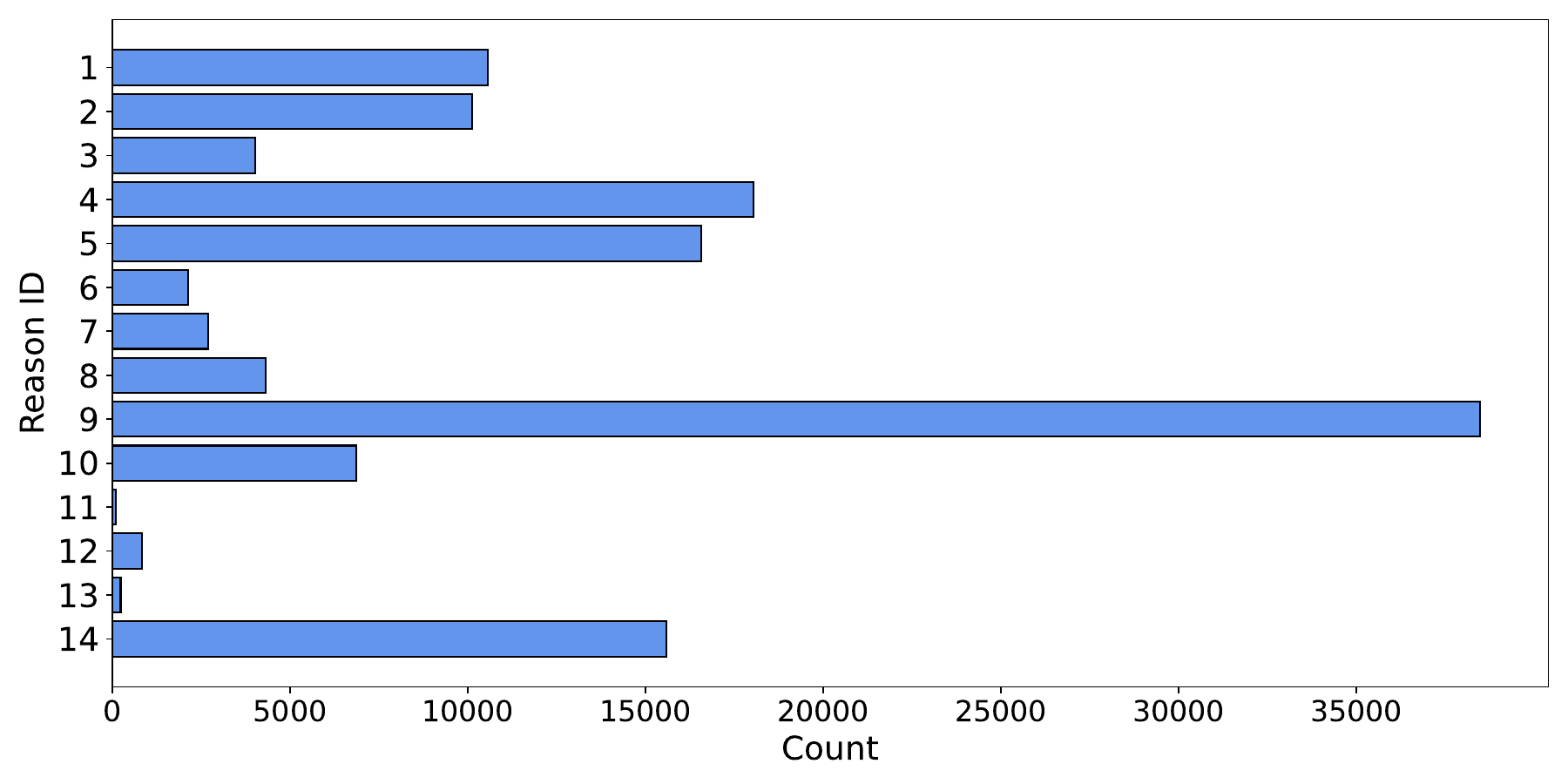}
  \caption{Distribution of obtained reasons from Listing~\ref{list:reasons}.}
  \label{fig:dist}
\end{figure}

\subsection{Ethics}
The study and annotation procedure received Institutional Review Board approval. Annotators were informed about the project goals and agreed to participate under these conditions. No personally identifiable information (e.g., age, race, gender) was collected. Instead, anonymized identifiers were used. The annotation platform preserved participant anonymity, and annotators were compensated per sample, corresponding to approximately \$8--10 per hour. Annotators were required to be native speakers of either English or Russian and preferably proficient in the other language. No individual was obligated to annotate.

\subsection{Instructions}

Unlike~\cite{warren2024better}, where the questionnaire was designed to avoid training annotators, we provided detailed instructions, training examples, and feedback on prediction accuracy during labeling, including additional training or removal of underperforming annotators. However, annotators were not guided on how to write explanations and were not shown the correct labels. Audio samples were presented in random order. For each audio sample, annotators answered the question: \texttt{Assess whether the audio sample contains original (genuine) human speech or synthesized (artificial) speech}. If ``genuine'' was selected, annotators provided a brief explanation. If ``artificial'' was selected, they provided a free-form explanation and selected relevant reasons from a predefined list of $14$ options, as shown in Listing~\ref{list:reasons}. 

\begin{lstlisting}[
    basicstyle=\footnotesize\ttfamily,
    numbers=left,
    numberstyle=\tiny,
    xleftmargin=1.2em,
    frame=none,
    breaklines=true,
    breakatwhitespace=true, caption={Annotation reasons list.}, label={list:reasons},
]
yt (1) Lack of fluency or coherence
(2) Unnatural pauses
(3) Uniform pauses between words throughout the audio
(4) Unusual intonation patterns
(5) Insufficient variation in speaking style
(6) Incorrect stress in common words
(7) Mispronunciation of common words
(8) Unusual or inconsistent accent
(9) Atypical voice characteristics
(10) Excessively fast speech
(11) Incorrect reading of abbreviations
(12) Verbalization of typographical errors
(13) Word-by-word repetition in cases of tautology
(14) Other (please specify)
Optional comment:
______________________________
\end{lstlisting}

Annotators were additionally provided with the following recommendations:
\begin{itemize}
    \item If you selected ``synthesized'' and indicated specific factors from the list,
please describe in detail where and how these cues manifested in the audio.
\item  If you are uncertain, explain which aspects appeared suspicious and which
seemed natural.
\item If you selected ``genuine'', describe the characteristics that led you to
perceive the speech as natural (e.g., realistic pauses, natural intonation,
breathing patterns, expressive variability).
\item Responses should be detailed and well justified. Short statements such as
``it sounds normal'' are not acceptable.
\end{itemize}

\begin{table}[th]
\caption{Evaluation results on \texttt{Test-1-HL}.}
\label{tab:main-results}

\resizebox{\linewidth}{!}{%
\begin{tabular}{lllll}
\toprule
\textbf{Model} &
\textbf{\makecell[l]{Train\\Set}} &
\textbf{Accuracy} &
\textbf{\makecell[l]{Balanced\\Accuracy}} &
\textbf{F1} \\
\midrule

\multirow{4}{*}{\makecell[l]{Wav2Vec2-\\AASIST}}
& Train-1-HL & 92.3 & 81.3 & 76.7 \\ \cline{2-5}

& Train-2-HL & 92.9 & 84.0 & 76.7 \\ \cline{2-5}

& \makecell[l]{Train-1-HL\\Train-2-HL}
& 92.3 & 83.9 & 76.6 \\
\midrule
\multirow{9}{*}{SALMONN-7b}
& Train-1-HL & 93.4 & 89.3 & 84.5 \\ \cline{2-5}

& Train-2-HL & 94.5 & 88.6 & 85.7 \\ \cline{2-5}

& \makecell[l]{Train-1-HL\\Train-2-HL}
& 93.1 & 83.8 & 80.4 \\ \cline{2-5}

& Train-2-R & 92.9 & 86.7 & 81.4 \\ \cline{2-5}

& \makecell[l]{Train-1-HL\\Train-2-R}
& 92.7 & 86.5 & 81.2 \\ \cline{2-5}

& \makecell[l]{Train-2-R\\Val-1-GRPO}
& 93.6 & 89.6 & 85.0 \\
\bottomrule
\end{tabular}%
}
\end{table}


\begin{table*}[t]
\caption{Reasoning samples for SALMONN-7B tuned on \texttt{Train-2-R} and evaluated on \texttt{Test-2-R}.}
\label{tab:test_reasoning}
\resizebox{\textwidth}{!}{%
\centering
\footnotesize
\setlength{\tabcolsep}{4pt}
\renewcommand{\arraystretch}{1.15}

\begin{tabular}{p{20em}c c p{20em}c c}
\toprule
\textbf{Human Comment} &
\textbf{Tags} &
\textbf{Label} &
\textbf{Model Reasoning} &
\textbf{Pred. Tags} &
\textbf{Pred. Label} \\
\midrule

I believe the audio is genuine. I heard a non-robotic voice. The speech pace varies, speeding up and slowing down. There is a natural rhythm following the meaning, clear pronunciation, slight tone changes, and audible breaths without artifacts.  
& -- & Real
& I believe the audio is genuine. The voice is non-robotic and the pace varies naturally. The rhythm sounds lively and human, with logical stress and clear pronunciation.  
& -- & Real \\

\midrule

The synthesized speech has high quality and clear intelligibility without artifacts. The voice is soft and neutral, but the speech rate is slightly fast and emotional engagement is limited.  
& (5),(10) & Fake
& The speech is too fast and uninterrupted.  
& (1),(2),(10) & Fake \\

\midrule

The voice is monotone and the narrator reads the sentence without pausing.  
& (1),(9) & Fake
& The speech is unnaturally fast; it is difficult for a person to speak this quickly without pauses or mistakes. Odd pronunciation of ``Europe''.  
& (1),(2),(9) & Fake \\

\midrule

Monotone speech. It's not entirely clear from the context what intonation was originally intended.  
& (4),(9) & Fake
& The voice sounds robotic, and the speech is unnatural and staccato.
& (4),(9) & Fake \\

\midrule

A genuine recording for a voice assistant. Environmental noise is audible (TV in the background).  
& -- & Real
& The speech sounds authentic: natural intonation, individual pronunciation of sounds, consistent speaking pace, and background noise consistent with a television.  
& -- & Real \\

\bottomrule
\end{tabular}
}
\end{table*}

\subsection{Annotations filtering and postprocessing}

For data filtering, the following strategy was applied. First, we excluded annotators with classification accuracy below $75\%$. Second, samples with incorrect class predictions were removed. Finally, we manually inspected 30–50 annotations per annotator. Considering their overall accuracy, we categorized responses as ''high'' quality ($\ge$ 85\% accuracy or more) or ''medium'' quality ( $\ge$ 75\% accuracy or more). After filtering, the dataset contained $124, 410$ annotations of $41,414$ audio samples ($32,045$ spoof and $9,369$ bona fide) provided by $37$ annotators. The distribution of selected reasons is shown in Fig.~\ref{fig:dist}. In total, $120,258$ non-empty comments were collected with an average length of $12$ words. Human annotations and questionnaire selections were translated to English and post-processed using Qwen-32B to produce reasoning-like traces. For a fidelity check, a random set of $100$ audios was evaluated to verify that the rewritten traces align with genuine human annotations.

\subsection{Splits}

The hard-label dataset (without reasoning)  was collected from the sources described in subsection~\ref{sec:audio_sources}, and is denoted as \texttt{Train-1-HL}. \texttt{Val-1-HL} is a whole ASVspoof~5 development subset. For \texttt{Test-1-HL}, we randomly selected $20,000$ samples from the ASVspoof~5 evaluation  subset, collecting $15,943$ spoof and $4,057$ bona fide audio. The reasoning dataset was split into train, validation, and test subsets of 114k, 8k, and 1k samples, respectively. These splits can be used for binary classification or CoT training and evaluation, e.g., \texttt{Train-2-HL} and \texttt{Train-2-R}.
We highlight that ASVspoof 5 splits do not share the same domain and have different non-overlapping generators, and comprise more modern generators than in WaveFake, FakeAVCeleb, or ASVspoof2021 datasets. 

\section{Methodology}

\subsection{Binary classification and CoT SFT training}

Despite strong overall performance, LLMs and LALMs often show weak zero-shot and few-shot results on new tasks, including SDD~\cite{gu2025allm4add}. Following this observation, we begin with SFT using LoRA applied to all linear projections of the LLM backbone. We use SALMONN~\cite{tang2023salmonn}, which combines Whisper~\cite{radford2023robust} and BEATS~\cite{chen2022beats} audio encoders with a Q-Former~\cite{li2023blip} adapter connected to the Vicuna-7B/13B~\cite{vicuna2023} LLM. The model receives concatenated audio features and textual instructions and is trained with cross-entropy loss on completion tokens only (prompt tokens are masked). 
For comparison, we also evaluate a strong conventional Wav2Vec2-AASIST~\cite{tak2022automatic} SDD model.
For \textbf{hard-label SFT}, the model is trained on the \texttt{Train-1-HL}, \texttt{Train-2-HL}, or their combination to output a binary answer (``\texttt{Final Answer: Real}'' or ``\texttt{Final Answer: Fake}''), {resembling ALLM4ADD's approach.} For \textbf{CoT SFT}, the model is trained on the \texttt{Train-2-R} to produce structured output in the following format:
\begin{quote}
  \texttt{<think>...</think>}\\
  \texttt{<reasons>[...]</reasons>}\\
  \texttt{<answer>Real/Fake</answer>},
\end{quote}
where \texttt{<think>} holds the free-form reasoning trace, \texttt{<reasons>} lists detected cues from the annotation taxonomy (Listing~\ref{list:reasons}), and \texttt{<answer>} gives the final binary prediction.

Two initialization strategies are explored: (i)~fine-tuning from the base (vanilla) LALM, and (ii)~fine-tuning from a hard-label SFT checkpoint. In both cases, the model is trained to generate the full structured output, including the reasoning trace, the list of detected artifact categories, and the binary prediction.  {Incorrectly annotated samples were excluded from CoT due to the unreliability.}

SALMONN-7B was fine-tuned using LoRA with rank $128$, $\alpha=256$, and dropout $0.2$. The learning rate started at $10^{-6}$ with a $5000$-step warm-up, reached $10^{-4}$, and decayed to $10^{-5}$. Audio was cropped or padded to 10s. GRPO weights were set to 0.2/0.7/0.1 for format/class/judge predictions, respectively.

\subsection{GRPO and grounding}
The LALM can generate fluent and structurally valid CoT traces that nevertheless do not correspond to the actual audio content and can be based on hallucinations or training CoT memorization. To address this issue,  we use an audio grounding stage to explicitly encourage the model to anchor its \texttt{<think>} traces and its \texttt{<reasons>} tags to perceptible acoustic evidence, such as Gaussian noise, time masking, and gain adjustments embedded deterministically in the audio signal. To further improve the quality and diversity of CoT, a GRPO~\cite{shao2024deepseekmath} was applied. Reward functions encouraged the model's correctness, following the format (\texttt{<think>}, \texttt{<reasons>}, \texttt{<answer>} tags) and adjusting to human preferences. For the latter, Qwen2.5-32B was requested to evaluate each generation on its coverage, relevance, logic (coherence with binary prediction) and helpfulness from $0$ to $10$. For each sample, we generated $6$ distinct reasoning traces to calculate rewards.

\subsection{Metrics}
For the primary metrics of SDD classification, accuracy, balanced accuracy, and F1 (positive class is bona fide) were considered.
Reasoning quality was evaluated using the GPT-5.1 \cite{gpt51} model over the same criteria set that were used in GRPO: recall, relevance, logic, helpfulness, with scores from 0 to 10. Sets of reasons (tags) were evaluated using the Jaccard similarity index. For predicted tags $\hat{Y}$ and ground-truth tags $Y$, it is defined as $J(Y,\hat{Y}) = \frac{|Y \cap \hat{Y}|}{|Y \cup \hat{Y}|}.$

\section{Results and discussion}
SALMONN-7B outperformed conventional Wav2Vec2-AASIST in terms of binary classification metrics, as shown in Table~\ref{tab:main-results}. Moreover, the obtained models demonstrate similar or better results compared to the open-source models, presented on Speech DF-Arena leaderboard~\cite{dowerah2026speech} .


Nonetheless, LALMs can demonstrate weak training stability and a drastic overfitting tendency; to mitigate this issue, the number of training iterations and learning rate might be decreased, while dropout and augmentation range are increased. 
Training on the combined dataset of \texttt{Train-1-HL} and \texttt{Train-2-HL} does not improve the results, compared to the separately trained checkpoints. Regarding hard-label evaluation of reasoning checkpoints, while reasoning-only SFT underperformed to the hard-label checkpoints, GRPO tuning improves the results. 
The hypothesis that hard-label SFT can help the reasoning SFT has not been confirmed. 
We highlight the decent reasoning performance on a set of samples from \texttt{Test-2-R}, which is reflected in Table~\ref{tab:test_reasoning}.
One can notice that SALMONN provides reasonable audio-grounded cues, such as referring to particular words' pronunciation and to background noises. Among the other findings, it is noticeable that grounding does not significantly improve the classification metrics, but diversifies the chain-of-thought traces, while GRPO further improves the reasoning. After applying these techniques to the model, the diversity and informativeness of the answers it generates, as well as its classification metrics, increase (see Table \ref{tab:main-results}). Binary classification Test-1-HL EER is $13.2\%$ (for SALMONN-7B model) and $15.7\%$ (for W2V2-AASIST model). CoT final decisions are strongly conditioned on the preceding trace, making EER redundant.
Despite these results, the resulting reasoning models still struggle with modern high-fidelity synthesis systems that were not present in the training data. LALM's reasoning often describes deepfake audio as genuine speech. {Nonetheless, one can train a model to output ``no audible artifacts'' on genuine speech, treating bona fide as an abstention class.}

Regarding CoT quality evaluation, GRPO does not yield a significant improvement in the Jaccard index. 
Namely, for the same audio sample, inter-annotator agreement on selecting each individual reason ranged from $0.05$ to $0.56$ in Fleiss’ kappa. However, the average pairwise Jaccard similarity between annotators’ selected reason sets was $0.78$ on the test set, which might serve as a reference point for the comparison with models' performance.
Given the noise in reasoning-tag markup, the scores are effectively unchanged: $0.6468$ for the GRPO checkpoint vs. $0.6264$ for the SFT checkpoint.

In an LLM-as-a-judge evaluation (five runs on $1,000$ reasoning samples), GRPO shows a small but non-significant improvement in reasoning-trace quality: mean score $5.74 \pm 1.49$ vs. $5.12 \pm 1.47$ for SFT. However, the observations suggest that without grounding and a GRPO, reasoning becomes poorer, as it lacks more audio cues.
We note that the judge's rubric is intentionally strict, which keeps absolute scores relatively low. 


\section{Conclusion}
This work addressed the problem of limited generalization and poor interpretability in modern speech deepfake detection systems. We introduced a novel human-annotated reasoning dataset for the SDD and proposed a HIR-SDD, a framework that enables models not only to perform binary spoof detection but also to provide human-interpretable explanations grounded in perceptual cues. Experimental results demonstrate that the proposed framework achieves competitive detection performance while producing meaningful reasoning traces aligned with human annotations. These findings suggest that the incorporation of proposed human-inspired reasoning traces can improve both transparency and reliability of SDD systems.
Future work includes improving robustness to unseen generative models and domain shifts, as well as refining the stability, quality, and evaluation of reasoning traces produced by LALMs.

\section{Generative AI use disclosure}
AI models were used only for grammar correction and for text refining.

\section{Acknowledgment}
This work was supported in part by a grant for research centers in artificial intelligence provided by the Ministry of Economic Development of the Russian Federation under agreement No. 000000C313925P4F0002, and by the Russian Science Foundation (RSF), grant No. 25-41-00091.


\bibliographystyle{IEEEtran}
\bibliography{mybib}

\end{document}